\documentclass[a4paper,fleqn,usenatbib,useAMS]{mnras}
\usepackage{graphicx}
\usepackage{amssymb}
\usepackage{array}
\usepackage{color}
\usepackage{longtable}
\usepackage{natbib}
\bibpunct{(}{)}{;}{a}{,}{,}

\newcommand{\degree}{\ensuremath{^\circ}}

\title[Magnetic fields in BRCs]{Magnetic fields in multiple bright-rimmed clouds in different directions of H$~$II region IC\,1396 - II}
\author[Soam et.~al]{Archana Soam$^{1,3}$\thanks{email:{archanasoam.bhu@gmail.com, archana@kasi.re.kr}},
G. Maheswar$^{2,3}$,
Chang Won Lee$^{1,4}$,
Neha S.$^{3,5}$,
Kee-Tae Kim$^{1}$
\\
$^{1}$ Korea Astronomy $\&$ Space Science Institute (KASI), 776 Daedeokdae-ro, Yuseong-gu, Daejeon 305-348, Republic of Korea.\\
$^{2}$ Indian Institute of Astrophysics, Kormangala (IIA), Bangalore 560034, India.\\
$^{3}$ Aryabhatta Research Institute of Observational Sciences (ARIES), Nainital 263129, India.\\
$^{4}$ University of Science \& Technology, 176 Gajeong-dong, Yuseong-gu, Daejeon, Republic of Korea\\
$^{5}$ Pt. Ravishankar Shukla University, Amanaka G.E.Road, Raipur, Chhatisgarh, India - 492010\\
}

\begin{document}
\date{Accepted------}

\pagerange{\pageref{firstpage}--\pageref{lastpage}} \pubyear{--}

\maketitle

\label{firstpage}

\begin{abstract}

Bright-­rimmed clouds form on the edges of H II regions affected by the high energy radiation from a central ionizing source. The UV radiation from the ionizing source results in compression and ionization causing either cloud disruption or further star formation. In this work, we present R$-$­band polarization measurements towards four bright-­rimmed clouds, IC1396A, BRC \,37, BRC\,38, and BRC\,39, located in the different directions of the H$~$II region, Sh­2-131, in order to map magnetic fields (B-­fields) in the plane of the sky. These BRCs are illuminated by the O star HD\,206267 and present a range of projected on ­sky geometries. This provides an opportunity to understand the magnetized evolution of BRCs. The B-­field geometries of the clouds deduced from the polarization data, after correction for foreground ISM contamination, are seen to be connected to the ambient B-­fields on the large scale. They seem to play an important role in shaping the cloud IC1396A and BRC\,37. BRCs\,38 and 39 show a broader and snubber head morphology possibly due to the B-­fields being aligned
with incoming radiation as explained in the simulations. A good general agreement is noted on comparing our observational results with the simulations supporting the importance of B-­fields in BRC evolution. This work is the first step towards
systematic mapping the B-­fields morphology in multiple BRCs in an expanding H$~$II region, extending the work presented by \citet{2017MNRAS.465..559S}.

\end{abstract}


\section{Introduction}

The H$~$II regions are formed when a massive star photoionizes and thus expand the surrounding medium. The formation of over-pressure ionozed bubbles are formed by the massive stars (OB-type) emitting the high energy photons (with rates of $10^{47}-10^{50}$~s$^{-1}$). The shocks driven from the ionizing source results into the expansion of these bubbles. When the shock is isothermal, a thin, dense and generally unstable gas and dust shell forms at the H$~$II region boundary. These boundaries are often found with peculiar and highly irregular structures. According to their structural appearances, these objects are named as fingers or pillars, speck globules, bright-rimmed clouds (BRCs), cometary globules (CGs) and elephant trunks \citep{1985prpl.conf..104L}. It was unclear whether these irregular structures are formed as a result of flow instabilities \citep{1954ApJ...120....1S, 1979ApJ...233..280G, 1996ApJ...469..171G, 1999MNRAS.310..789W} or preexisting dense structures from the ambient interstellar medium advected through the ionization front \citep{1983A&A...117..183R}. But recently \citet{2014MNRAS.444.1221K, 2015MNRAS.450.1017K} have systematically  investigated these objects and revealed that an initially ellipsoidal and inclined molecular cloud to an ionizing source could form most of the observed irregular structures from a filament and asymmetrical BRCs to a horse-head etc, by changing the initial density, geometry, inclination angle and the strength of the ionization flux. This work, combined with the previous investigation by \citet{2009ApJ...692..382M} on the formation mechanism of symmetrical type A, B and C type BRCs, provided a completed set mechanisms for the formation of various structures found on the boundaries of H$~$II regions.

The radiation from the massive stars drives an implosion into the globules sitting in the vicinity. The radiative driven implosion (RDI) mode presupposes a rather structured ambient medium to which the H$~$II region expands rapidly in directions where the density is relatively low. As this Rapid-type (R-type) ionization front encounters a relatively high density globule, it translates to a Delayed-type (D-type) and drives a convergent shock into the globule. Maximum compression of the globule occurs in this initial phase causing it to implode. Existing simulations of RDI focused on the onset and the efficiency of triggered star formation in isolated pre-existing clumps \citep{2011ApJ...736..142B, 2012MNRAS.426..203H}, and on the driving of turbulence. The accretion luminosity of the embedded protostar in the globules could be increased as a result of the enhanced accretion rate due to strong compression during RDI process. But, on the other hand the luminosity can be decreased in the RDI process due to the photoevaporation of the parent core causing a decrease in the core mass \citep{2007A&A...467..657M}. For investigating this process in detail and to model the collapse of the cores exposed to ultraviolet (UV) radiation from massive stars, \citet{2007A&A...467..657M} presented some numerical simulations and estimated the mass loss rate dependence on the initial density profiles of cores and variation of UV fluxes. Their study  also derived the simple analytic estimates of the accretion rates and final masses of the protostars. 


First 3-D magnetohydrodynamical (MHD) simulations towards magnetized globules were presented by \citet{2009MNRAS.398..157H}. They found that photoevaporating globules will evolve into more flattened sheet like structures compared to the non-magnetic cases when the cloud initially has a strong B-field (i.e. 100 times the thermal pressure) perpendicular to the UV radiation direction. Later, they extended this work by adding B-fields of various strengths and orientations \citep{2011MNRAS.412.2079M}. One of the important results they obtained was that an initially perpendicular but weak and medium strength B-fields are finally swept away and get aligned with the pillar during its dynamical evolution. This is consistent with the observed field orientations in M16 \citep{2007PASJ...59..507S} and in some CGs \citep[e.g., ][]{1987ApJ...319..842H, 1996MNRAS.279.1191S}. However, in CG 30-31 (BRC 51), \citet{1999MNRAS.308...40B} found that the projected B-fields are oriented perpendicular to the tail of the globule. Above results obtained on CG 30 -31 are different from the previous polarization studies towards various H$~$II regions like RCW41 \citep{2012ApJ...751..138S}, NGC3576 \citep{2009ASPC..404...27S} and Sh 156 \citep{1983MNRAS.202...11K} where the B-fields are found to be parallel to the tail part. So far very few attempts have been made observationally to map the B-field geometry towards some BRCs such as BRC\,20 \citep{2011ApJ...743...54T}, BRC\,51 (CG 30-31, \citealt{1999MNRAS.308...40B}), BRC\,74 \citep{2015ApJ...798...60K} and BRC\,89 \citep{2014ApJ...783....1S}.

IC1396 is a young and active H$~$II regions in the Cep OB2 containing clustered OB stars \citep{1991ApJ...370..263S}. It appears that the expansion of this H$~$II region has resulted into sweeping up a molecular ring of radius $\sim$12 pc \citep{1995ApJ...447..721P}. This region contains 15 small globules which are found embedded with IRAS sources \citep{1991ApJ...370..263S}. The large scale dynamical study using CO molecular line observations has been presented by \citet{1995ApJ...447..721P} and \citet{1996A&A...309..581W}. \citet{1991ApJS...77...59S} and \citet{2005A&A...432..575F} found a rich population of BRCs and CGs seen in silhouette against the emission nebulae residing on the large molecular shell surrounding the H$~$II region. The dominant source of UV radiation in H$~$II region IC1396 is a O6.5V star HD206267 \citep{1984ApJ...286..718W, 1995ApJ...447..721P}. This star is the member of Trumpler 37, a young open cluster which is located at the center of the Cep OB2 association \citep{1968ApJ...154..923S}. From the photometric observations by \citet{1976PASP...88..865G} and \citet{1968ApJ...154..923S}, \citet{1995ApJ...447..721P} found 12 stars with spectral types earlier than B1. Clearly the hottest star HD206267 (O6.5) in the center is the primary source of ionizing UV radiation.

To understand the effects of B-fields in the structural evolution of the clouds, it is important to the perform a systematic polarimetric study by selecting an H$~$II region containing multiple BRCs at different locations, with known properties of ionizing source(s). Relatively closer regions (with less foreground extinction) with less complex geometry would be easier to study systematically. Assuming a preferred orientation of B-field prior to the formation of a H$~$II region, the clouds in the different directions would show different magnetized evolution because of the different orientation of ionizing radiation. \citet{2017MNRAS.465..559S} have chosen Sh 185 region with IC 63 and IC 59 nebulae in different direction of the radiation from ionizing source $\gamma$ Cas but there were only two nebulae studied in that region. In the present study, we show the polarization results of four BRCs associated to IC1396 H$~$II region. We have chosen these BRCs in north (BRC\,38 or IC1396N), south (BRC\,37), east (BRC\,39) and west (IC1396A and BRC\,36) directions of this H$~$II region. 

This study will give us an insight into the magnetized evolution of the multiple BRCs associated to the H$~$II region as a function of the direction of the ionizing radiation. This paper presents the data acquisition information in section 2. Results of the polarimetric observations are shown in section 3. In section 4, we discuss and analyze the results obtained and further conclude our findings in sections 5.


\section{Data acquisition}\label{sec:obs}

The polarization measurements towards IC1396 HII region were obtained on ten nights from October 2013 to January 2014 (see Table \ref{tab:obslog}) using Aries IMaging POLarimeter \citep[AIMPOL;][]{2004BASI...32..159R} at 1 m optical telescope situated in Aryabhatta Research Institute of Observational Sciences (ARIES), India. The mean exposure time per observed frame per half wave plate (HWP) angle was $\sim$ 250 sec for obtaining good signal-to-noise ratio. We obtained seven frames on each position of the HWP in one field and repeated the procedure for covering the four BRCs in IC1396 region. The details of the instrument and the procedure of data reduction for obtaining polarization values from observed data has been given in our previous papers \citep{2013MNRAS.432.1502S, 2015A&A...573A..34S, 2017MNRAS.465..559S}. Care has been taken in removing the instrumental polarization from the measurements by considering a constant value observed in various studies using this instrument. This value has been reported as 0.1\%  \citep[see ][]{2013MNRAS.432.1502S, 2015A&A...573A..34S, 2016A&A...588A..45N}. We also observed the polarized standard stars (results shown in Table \ref{tab:std}) from \citet{1992AJ....104.1563S} to determine the reference direction of the polarizer. A good correlation has been noticed among the observed values and the standard values given in \citet{1992AJ....104.1563S}. These values were used to obtain the zero point offset correction which was applied later on the observed position angles of the target stars.

\begin{table}
\caption{Log of polarimeteric observations towards IC1396 region in ${\rm R_{kc}}$ filter ($\rm \lambda_{eff}$=0.760$\mu$m).}\label{tab:obslog}
\begin{tabular}{p{1.3cm}p{6cm}}\hline
 Cloud ID          &  Date of observations (year, month,date)\\
\hline
 IC\,1396           & 2013, Oct, 29; 2013, Nov 08, 09, 10, 26, 28 \\
                    &  2013, Dec, 01, 03; 2014, Jan 03, 05 \\
\hline
\end{tabular}
\end{table}

\begin{table}
\caption{Results of observed polarized standard stars.}\label{tab:std}
\begin{tabular}{lll}\hline
Date of     &P $\pm$ $\sigma_P$ 	&  $\theta$ $\pm$ $\sigma_{\theta}$  \\
Obs.		&(\%)            		& ($\degree$)                           \\\hline
\multicolumn{3}{l}{{\bf HD 236633}}\\
\multicolumn{3}{l}{$^\dagger$Standard values: 5.38 $\pm$ 0.02\%, 93.04 $\pm$ 0.15$\degree$}\\
01 Dec 2013 & 5.5 $\pm$ 0.1     & 92 $\pm$ 1\\
03 Dec 2013 & 5.6 $\pm$ 0.1     & 93 $\pm$ 1 \\
03 Jan 2014	& 4.8 $\pm$ 0.2     & 92 $\pm$ 1 \\
05 Jan 2014	& 5.3 $\pm$ 0.1     & 92 $\pm$ 1 \\\hline
\multicolumn{3}{l}{{\bf HD 25443 }}\\
\multicolumn{3}{l}{$^\dagger$Standard values: 4.73 $\pm$ 0.05\%, 133.65$\pm$ 0.28$\degree$}\\
08 Nov 2013 & 4.9 $\pm$ 0.1     & 133 $\pm$ 2 \\
09 Nov 2013 & 4.9 $\pm$ 0.1     & 132 $\pm$ 1  \\ 
03 Dec 2013 & 4.8 $\pm$ 0.1     & 133 $\pm$ 1  \\ \hline
\multicolumn{3}{l}{{\bf BD$+$64$\degree$106}}\\
\multicolumn{3}{l}{$^\dagger$Standard values: 5.69 $\pm$ 0.04\%, 96.63 $\pm$ 0.18$\degree$}\\
01 Dec 2013 & 5.3 $\pm$ 0.1     & 96 $\pm$ 1   \\\hline
\multicolumn{3}{l}{{\bf HD 19820}}\\
\multicolumn{3}{l}{$^\dagger$Standard values: 4.53 $\pm$ 0.02\%, 114.46$\pm$ 0.16$\degree$}\\
29 Oct 2013 & 4.4 $\pm$ 0.1     & 114 $\pm$ 1  \\
03 Dec 2013 & 4.5 $\pm$ 0.1     & 113 $\pm$ 1  \\\hline
\multicolumn{3}{l}{{\bf HD204827}}\\
\multicolumn{3}{l}{$^\dagger$Standard values: 4.89 $\pm$ 0.03\%, 59.10 $\pm$ 0.17$\degree$}\\
29 Oct 2013 & 5.0 $\pm$ 0.2 	& 61 $\pm$ 7 \\        \hline
\multicolumn{3}{l}{{\bf HD43384}}\\
\multicolumn{3}{l}{$^\ddagger$Standard values: 4.89 $\pm$ 0.03\%, 59.10 $\pm$ 0.17$\degree$}\\
28 Nov 2013 & 4.7 $\pm$ 0.1  & 57 $\pm$ 3  \\   \hline
\end{tabular}

$^\dagger$ In $R_{c}$ band from \citet{1992AJ....104.1563S} \\
$\ddagger$ Values obtained from \citep{1982ApJ...262..732H}\\
\end{table}

\section{results}\label{reslt}
\begin{figure*}
\centering
\resizebox{16cm}{14cm}{\includegraphics{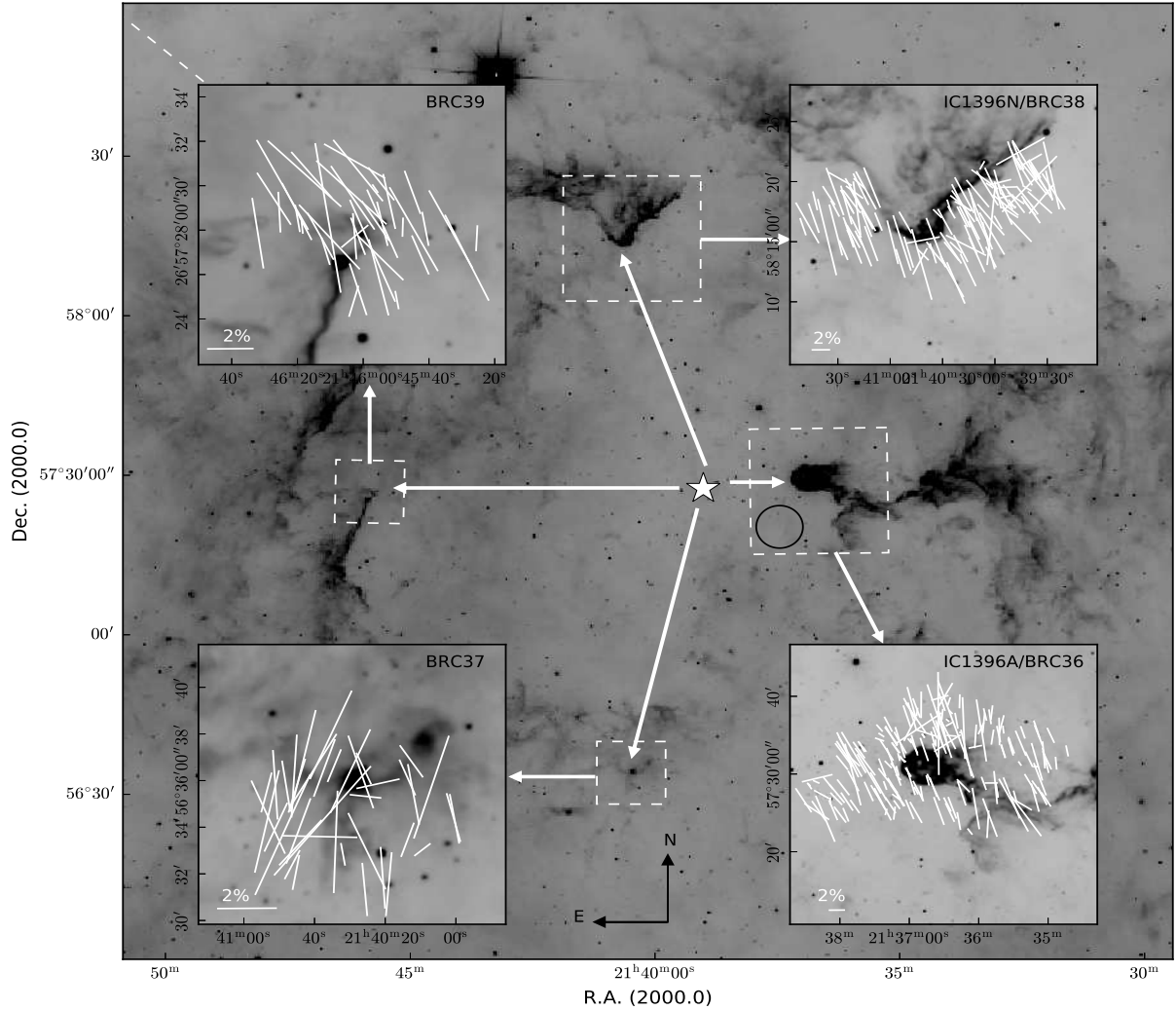}}
\caption{Optical polarization vectors over-plotted on WISE 12$\mu$m image of IC1396 H$~$II region. The structures IC1396\,A, BRC\,37, BRC\,38 and BRC\,39 with the B-field geometry are shown in zoomed-in images. Line segments with 2\% polarization are shown for scaling. A white dashed line shows the position angles of the Galactic plane. The black circle in the dashed panel on the center right fo the image shows the field of view of the instrument (AIMPOL). Position angles are measures from north increasing towards east. The north and east directions are shown in the figure.}\label{Fig:polwise}
\end{figure*}

\begin{figure*}
\centering
\resizebox{12cm}{12cm}{\includegraphics{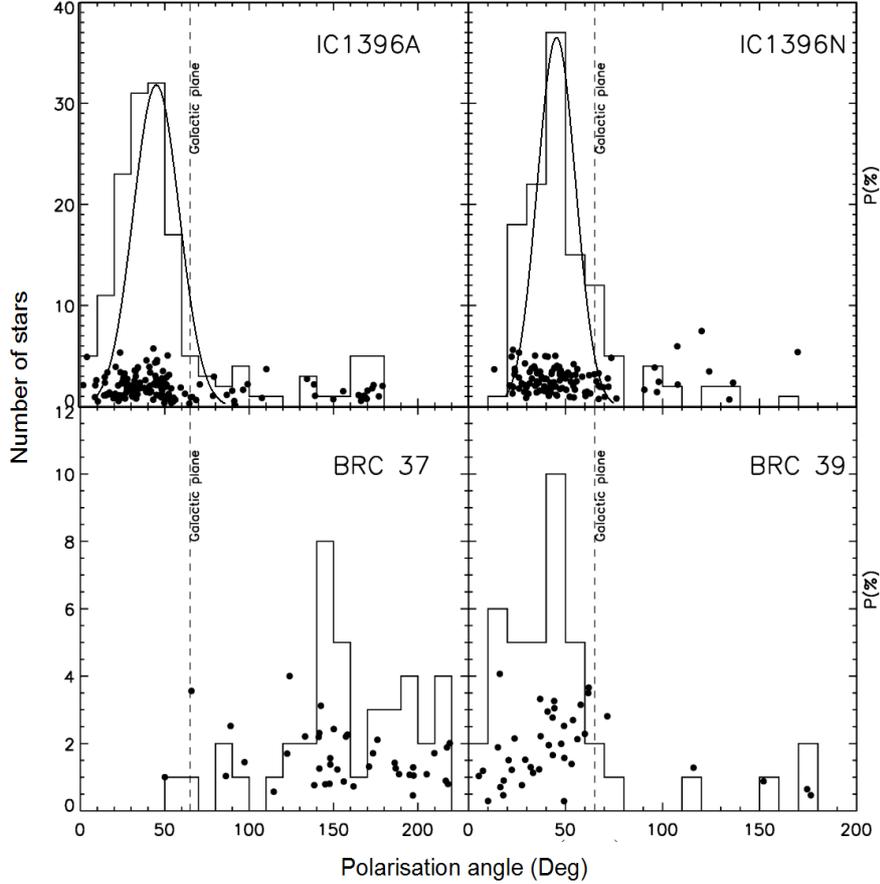}}       
\caption{Plots of the P versus $\rm \theta_{P}$ of stars excluding peculiar stars projected on IC\,1396A, BRC\,37, BRC\,38 (IC\,1396N), and BRC\,39. The histograms of the $\rm \theta_{P}$ with binsize 10$^\circ$ are also presented. The position angles of the Galactic plane at the latitude of the clouds are shown using dashed linse.}\label{Fig:P_PA_hist}
\end{figure*}

The results of our optical polarization measurements towards the stars projected on the fields of IC1396A (a.k.a. to elephant trunk nebula), BRC\,37, BRC\,38 (a.k.a. IC1396N) and BRC\,39 are presented in this work. The vectors corresponding to 374 polarization measurements, are overlaid on  WISE\footnote{WISE image courtesy of NASA, JPL/Caltech, WISE Team} 12 $\rm \mu$m image containing these fields in Fig. \ref{Fig:polwise}. The expanding circular H$~$II region IC1396 is clearly perceptible in the image. The direction of ionizing radiation and the position of ionizing source are also shown. The mean values of polarization measurements towards four BRCs are given in Table \ref{tab:mean_pol}. The location of these clouds are shown in the zoomed images in north (BRC 38), south (BRC 37), east (BRC 39) and west (IC1396A). The orientation of Galactic plane is shown with dashed white line at a position angle of 65$\degree$. Black open circle on the image shows the field of view (8$\arcmin$ diameter) of the instrument (AIMPOL) used for the observations. In our study we mapped the B-field geometry of the outer low density parts of the clouds. The distribution of degree of polarization (P) and position angle ($\rm \theta_{P}$) towards the four BRCs observed in IC1396 are shown in Fig. \ref{Fig:P_PA_hist}. We segregated the known YSOs, emission line sources, known binary systems and the normal stars from our observed sample based on the information from Simbad and literature. YSOs are identified in the vicinity of IC1396A \citep{2005AJ....130..188S, 2006AJ....132.2135S}, BRC\,37, BRC\,38 \citep{2010ApJ...717.1067C} and BRC\,39.


\begin{table}
\caption{Mean values of polarization results obtained towards BRCs studied in this work.}\label{tab:mean_pol}
\scriptsize
\begin{tabular}{lllr}\hline
Object  & No. of stars & $\rm P\pm\sigma_{P}$& $\rm \theta\pm\sigma_{\theta}$ \\
& & (\%) & ($^{\degree}$) \\\hline
IC1396A &173&2.0$\pm$1.1 &  53$\pm$18 \\
BRC\,37 &40&1.6$\pm$0.8 &  98$\pm$50 \\
BRC\,38 &121&2.7$\pm$1.2 &  51$\pm$17 \\
BRC\,39 &40&1.8$\pm$1.0 &  50$\pm$21 \\ \hline
\end{tabular}
\end{table}


\subsection{Polarimetric results of the YSOs found in the sample}

YSOs are found to show intrinsic polarization which is attributed to the asymmetric distribution of circumstellar material either as a disk or a flattened envelope \citep{1972ApJ...175..127B, 2011Prama..77...19B}. Therefore, we separated the polarization measurements of YSOs from the target stars. Out of the observed sources towards IC1396, 19 stars are identified as YSOs in Simbad. Hence we have polarization measurements of 19 YSOs towards this region out of which seven stars are projected on IC1396A \citep{2005AJ....130..188S, 2006AJ....132.2135S} and twelve stars are located towards BRC\,38 \citep{2010ApJ...717.1067C}. The mean values of P and $\rm \theta_P$ with their standard deviations towards these YSOs are estimated to be $1.2\pm1.1$\% and 55$\pm44^\circ$ in IC1396A and $2.4\pm1.2$\% and 46$\pm24^\circ$ towards BRC\,38, respectively.


\section{Discussion}\label{Discussion}

\subsection{Distance of IC1396 H$~$II region}\label{dist}

IC1396 is found to be at $\sim750$ pc by \citet{1976PASP...88..865G} using photometric and spectroscopic observations. It is a large ($>$ 2 degree) and evolved  H$~$II region ionized by HD 206267 located near its center. HD206267 lies in Trumpler 37 cluster which is located in the Cep OB2 association at a distance of about 800 pc. \citet{2002AJ....124.1585C} estimated distance to Trumpler 37 as 870$\pm$80 pc. Majority of the studies towards this H$~$II region assume a distance of $\sim750$ pc to the clouds associated to this region. We have adopted a distance of $\sim750$ pc \citep{1976PASP...88..865G} to this H$~$II region for our further analysis.

We measure the polarization of stars background to the cloud for mapping the sky component of the B-fields in that cloud but it is very important to consider the polarization contribution of the foreground dust in the line of sight to obtain the intrinsic polarization solely from the dust in the cloud . This contribution can vary depending on the distance of the cloud. For example, this contribution has not been found changing the measured values significantly as seen  in various studies done on clouds located at distances less than 500 pc \citep[e.g., ][]{2009ApJ...704..891L, 2013MNRAS.432.1502S}. But it can affect the polarization measurements if there are relatively smaller P values ($\lesssim 1\%$) and the cloud distance is relatively larger ($\rm \gtrsim 500 pc$). Since the distance of the H$~$II region studied here is $\sim$750 pc, we decided to remove the foreground polarization component added to the results by observing some stars located foreground to the region. Fig. \ref{Fig:polwise} shows the uncorrected polarization vectors overlaid on IC1396 region.

\subsection{Subtraction of foreground polarization component}\label{FG_corr}

\begin{table}
\caption{Polarization values of observed stars foreground to IC1396 region.}\label{tab:fg}
\begin{tabular}{llllll}\hline
Id&Star	Name	&V   	& P $\pm$ $\epsilon_P$ & $\theta$ $\pm$ $\epsilon_{\theta}$  &D$^{\dagger}$\\ 
  &				&(mag)	& (\%) 			&($\degree$)	&(pc) \\\hline
1  & HD 207049          &8.9&  0.95$\pm$    0.08  &   5$\pm$    2 &  379\\
2  & HD 206081          &7.6&  1.07$\pm$    0.07  &  64$\pm$    1 &  446\\
3  &  HD 209744           & 6.7&   0.65 $\pm$   0.20&   55$\pm$   8&  450 \\
4  &  HD 210628           & 6.9&   1.26 $\pm$   0.20&   60$\pm$   4&  480 \\
5  & HD 239728          &8.8&  0.58$\pm$    0.06  & 115$\pm$    3 &  568\\
6  & HD206267A          &5.6&  1.02$\pm$    0.07  &  39$\pm$    2 &  606\\ \hline
\end{tabular}

$^{\dagger}$ The parallax measurements from Hipparcos \citep{2007A&A...474..653V} and GAIA Catalogues \citep[TGAS;][]{2016yCat.1337....0G} are used for the distance estimation.
\end{table}

\begin{figure}
\resizebox{8.0cm}{9.7cm}{\includegraphics{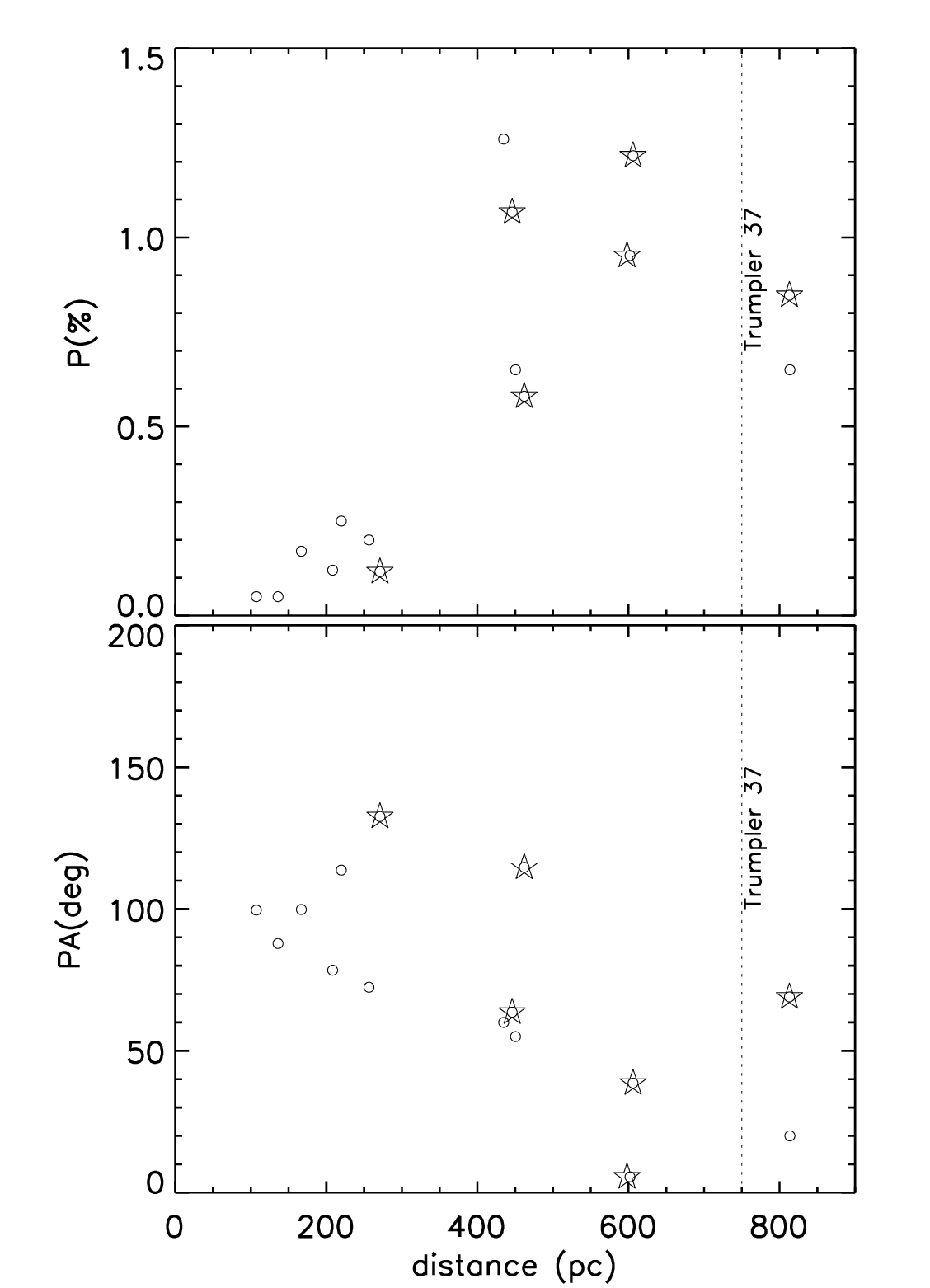}}
\caption{The change in the degree of polarization and the position angles with the distances of the foreground stars have been plotted in upper and lower panels, respectively. The open circles show the stars selected from Heiles catalogue \citep{2000AJ....119..923H}. The targets shown by stars are the foreground stars observed by us. The distance of these stars are taken from the Hipparcos \citep{2007A&A...474..653V} and GAIA Catalogues \citep[TGAS;][]{2016yCat.1337....0G}. The dotted line represents the distance of cluster Trumpler 37.}\label{Fig:fg_p_pa_dist}
\end{figure}

\begin{figure}
\resizebox{8.0cm}{10.2cm}{\includegraphics{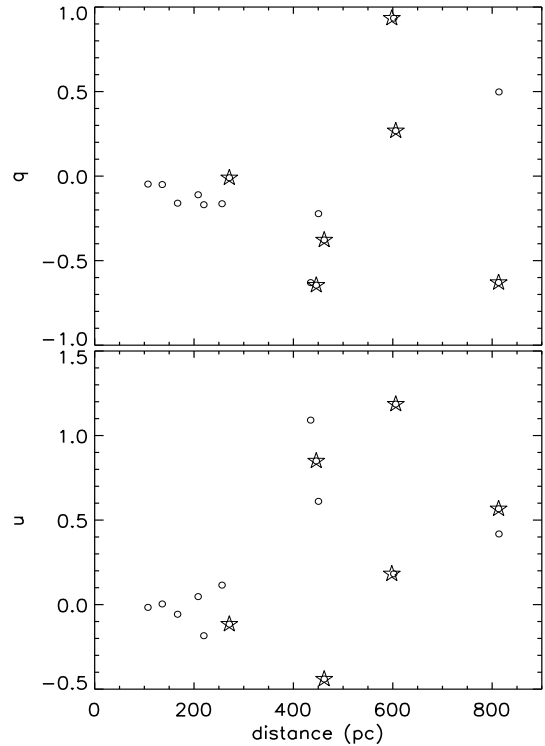}}
\caption{Same as Fig. \ref{Fig:fg_p_pa_dist} except for variations of Stokes parameters q (upper panel) and u (lower panel) with distances.}\label{Fig:fg_q_u_dist}
\end{figure}

The polarization caused by the material in ISM in a particular line of sight must be taken into account. Although we did not notice major change after the ISM contribution subtraction in the polarization results observed towards the cloud IC 63/59 \citep{2017MNRAS.465..559S} at a distance of $\sim$200 pc but we suspect that there may be a significant contribution of the ISM polarization in the results obtained towards IC1396 H$~$II region due to its far location ($\sim$750 pc). 

For subtracting the interstellar polarization component from measured values, we searched for the stars that are located within a radius of $1^\circ$ around the center of IC1396 H$~$II region with their available parallax measurements in Hipparcos \citep{2007A&A...474..653V} and GAIA\footnote{http://www.esa.int} Catalogues \citep[TGAS;][]{2016yCat.1337....0G}. We only selected the normal stars excluding the peculiar sources such as emission line stars, stars in a binary or multiple system as per the information given in Simbad. The stars with the values of ratio between the parallax measurements and their uncertainties as $\geq$ 2, have been considered for the analysis. Thus found six stars with their observed polarization results from this work, are listed in Table \ref{tab:fg} with the ascending order of their distances. We also selected normal stars in available in $1^\circ$ radius from Heiles catalogue \citep{2000AJ....119..923H} whose polarization values are available in the catalog. We have restricted these search of stars upto a distance of 800 pc. The variations of the degree of polarization and position angle corresponding to foreground stars with their distances are shown in Fig. \ref{Fig:fg_p_pa_dist}. The distribution of the Stokes parameters q and u with their distances is shown in Fig. \ref{Fig:fg_q_u_dist}. The dotted lines in the plot shows the distance of cluster Trumpler 37. In these figures, it is visible that there are two populations both in q and u which have different P and $\rm \theta$ values at distances $\leq$300 pc and greater than 300 pc. It might be possible that the some cloud at distance between 200-300 pc is responsible for the change in the polarization values beyond 300 pc. The stars within 300 pc to 650 pc are the immediate foreground to the clouds in IC1396 H$~$II region. Therefore, we have used the observed values of the stars within 300 pc to 650 pc for correcting the foreground contribution.

\begin{figure*}
\centering
\resizebox{16cm}{14cm}{\includegraphics{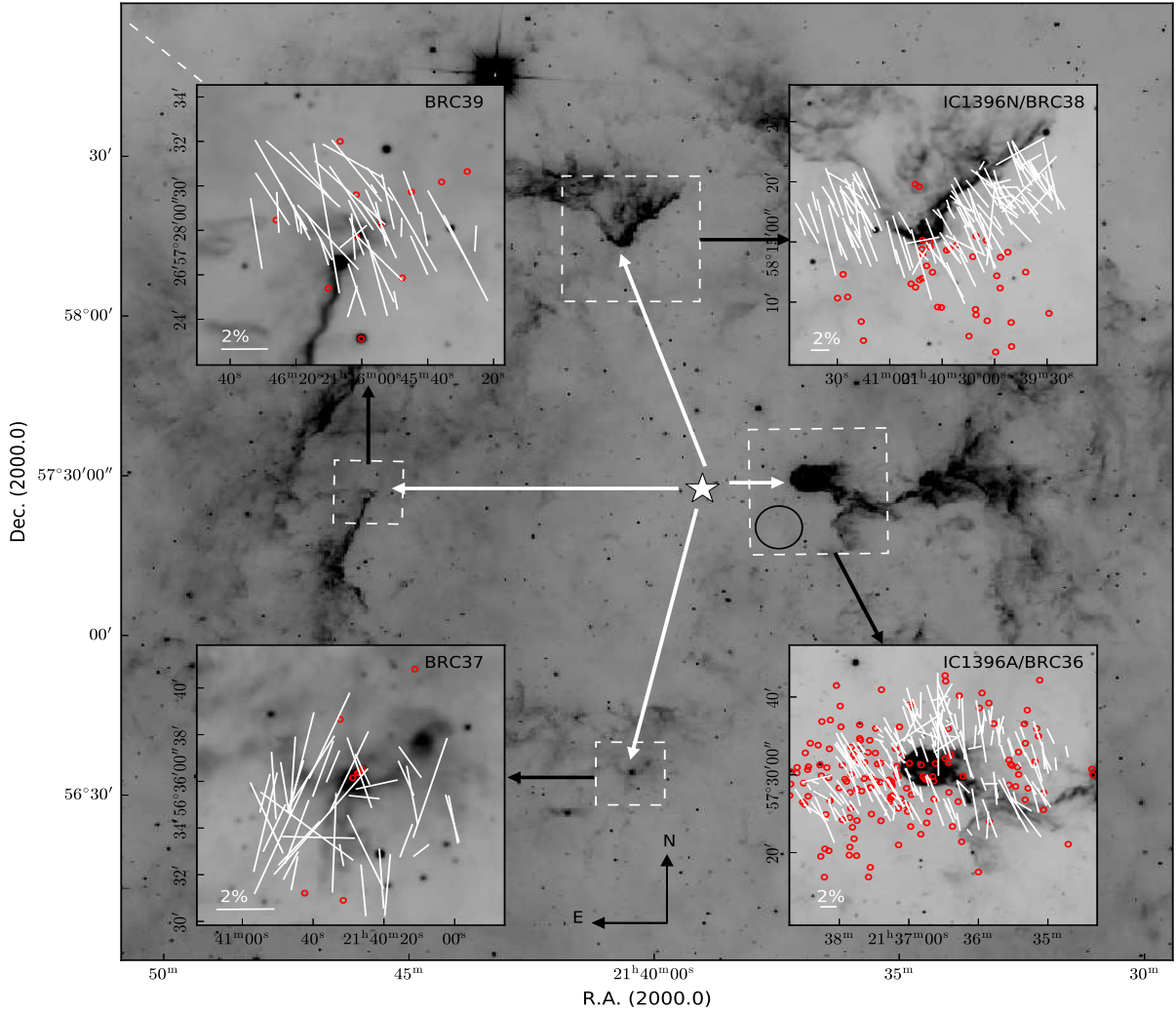}}
\caption{Polarization vectors (after subtracting the foreground polarization) overlaid on WISE 12$\rm \mu$m images of IC1396A, BRC\,37, BRC\,38 and BRC\,39. Line segemnts with 2$\%$ polarization are shown to scale the value of P. The dashed line shows the position angle of Galactic plane at the latitude of the cloud. The positions of YSOs ($\rm H_{\alpha}$ emission stars) distributed towards BRCs are shown using red open circles. Position angles are measures from north increasing eastward (directions shown in the figure).}\label{Fig:corr_pol}
\end{figure*}

To obtain the intrinsic polarization from the cloud dust only, we subtracted the polarization values of the foreground stars (shown in Table \ref{tab:fg}) from our measured values towards the clouds. For this purpose we followed the procedure with detailed discussion given in our previous papers \citep{2013MNRAS.432.1502S, 2017MNRAS.464.2403S}. We have overlaid the vectors with corrected polarization angles towards IC1396A, BRC\,38, BRC\,37 and BRC\,39 in Fig. \ref{Fig:corr_pol} and these values have been used for further analysis. 


\subsection{Magnetic field geometry}\label{MF_ic1396}

\begin{figure}
\resizebox{9cm}{8cm}{\includegraphics{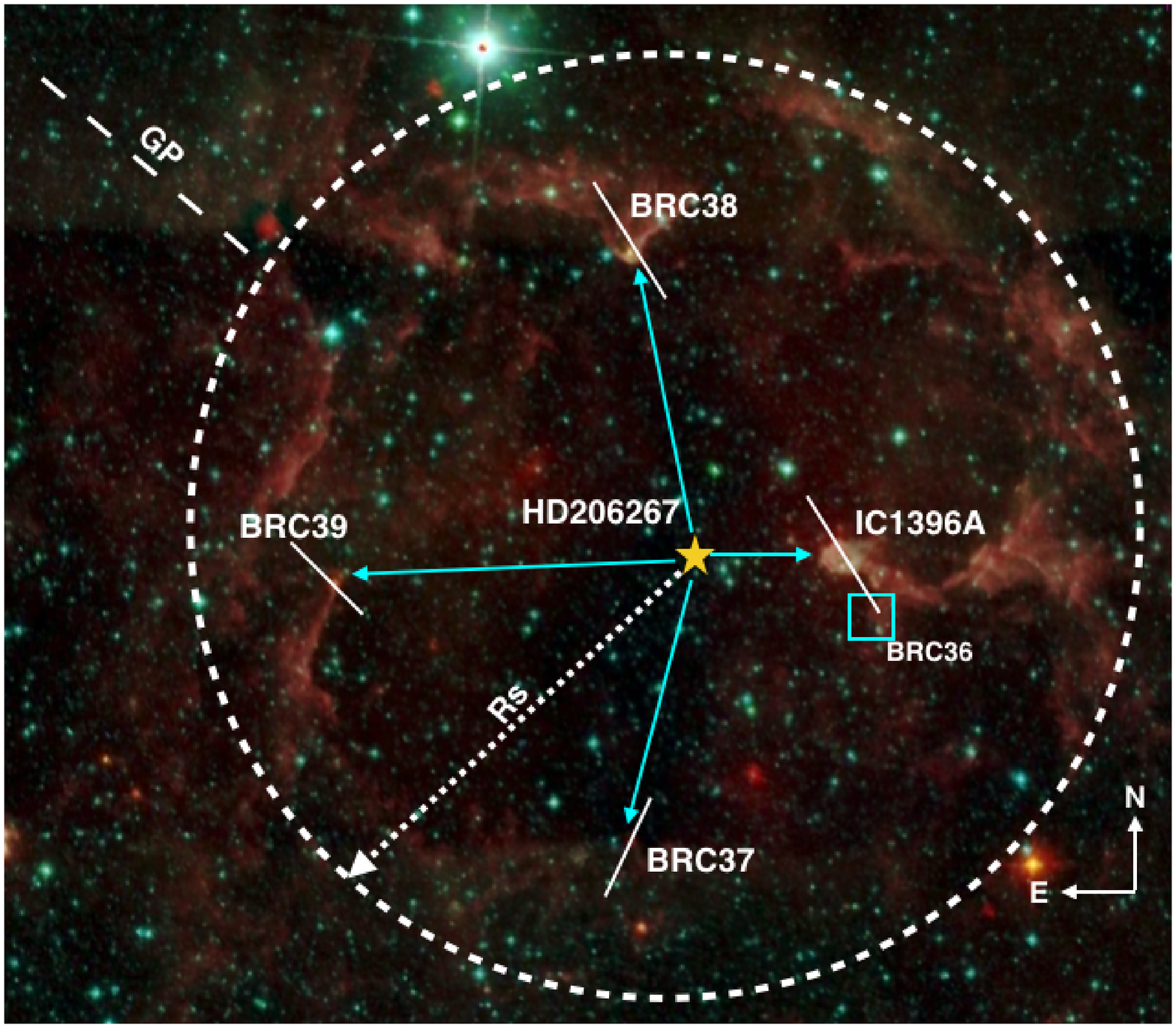}}
\caption{Figure shows the color composite image of IC1396 region made using WISE data. The image is labeled with the position of four BRCs in north, south, east and west. The position of the BRC 36 structure associated to elephant trunk nebula, is shown using a box in cyan color at the position of peak 850$\rm \mu$m peak emission \citep{2008A&A...477..557M}. The ionizing source HD206267 (spectral type O6.5V) is also located in the image. The white line segments plotted on the BRCs represent the mean B-field orientation in these BRCs. The Str$\rm \ddot{o}$mgren sphere \citep{1939ApJ....89..526S} created by ionizing radiation from HD206267 is shown by dashed circle. }\label{Fig:meanpol}
\end{figure}

We have carried out a systematic polarization study to map the B-fields in the H$~$II region IC1396 by observing the multiple BRCs associated to this region. Fig. \ref{Fig:meanpol} shows the color composite image of IC1396 region made using WISE data. The image is labeled with the position of four BRCs in north, south, east and west directions. The position of the BRC 36 towards IC1396A, is shown using a box in cyan color at the position coincident to the peak 850$\rm \mu$m peak emission \citep{2008A&A...477..557M}. The ionizing source HD206267 (spectral type O6.5V) is also located in the image. The directions of ionizing radiations towards BRCs are also shown. The white line segments plotted on the BRCs represent the mean B-field orientation. The Str$\rm \ddot{o}$mgren sphere \citep{1939ApJ....89..526S} created by ionizing radiation from HD206267 is shown by dashed circle. The mean value of the B-fields direction in these BRCs are inferred from the polarization vectors after correcting for the foreground component. We estimated the incident angles of the ionizing photons towards these BRCs by joining a line from the ionizing source HD 206267 radially to the positions of 850$\rm \mu$m dust emission peaks in these clouds identified by \citet{2008A&A...477..557M}. The directions of incident ionizing photons are estimated to be $90^\circ$ and $86^\circ$ in IC1396A and the globule BRC 36 located towards the tail of IC1396A, respectively. The angles of incident radiation towards BRC 37, BRC 38 and BRC 39 are found to be $\sim 160^\circ$, $\sim 30^\circ$ and $\sim 90^\circ$, respectively. Thus the projected offset angles between the incident ionizing radiation and the mean B-field orientation in BRC\,37, BRC\,38, IC1396\,A and BRC\,39 are estimated to be $\sim 10^\circ$, $\sim 20^\circ$, $\sim 35^\circ$ and $\sim 40^\circ$, respectively. These results imply that globally the B-fields are slanted w.r.t. the direction of the ionizing photons in the H$~$II region IC1396. In BRC 37, the least value of the offset between the B-field direction and the ionizing photons suggest that the ambient B-fields are almost (though not exactly) parallel to the ionizing radiation. The inclination of the B-fields and ionosing radiation in other three clouds are more than that in BRC\,37. These results can be compared to the 3-D MHD simulations published by \citet{2009MNRAS.398..157H} and \citet{2011MNRAS.412.2079M} where the B-fields in the globules slanted w.r.t. the ionizing radiation are considered. The B-fields in these BRCs associated to H$~$II region IC1396 (except BRC 37) are found to be parallel to the Galactic plane. This is in agreement with the results obtained by \citet{2005AAS...206.5401N, 2011Natur.479..499L} in the studies of B-field morphology towards GMCs. This is also noticed in the results towards BRCs associated to Sh 2-185 region explained in \citet{2017MNRAS.465..559S}.

\begin{figure*}
\centering
\resizebox{16cm}{10cm}{\includegraphics{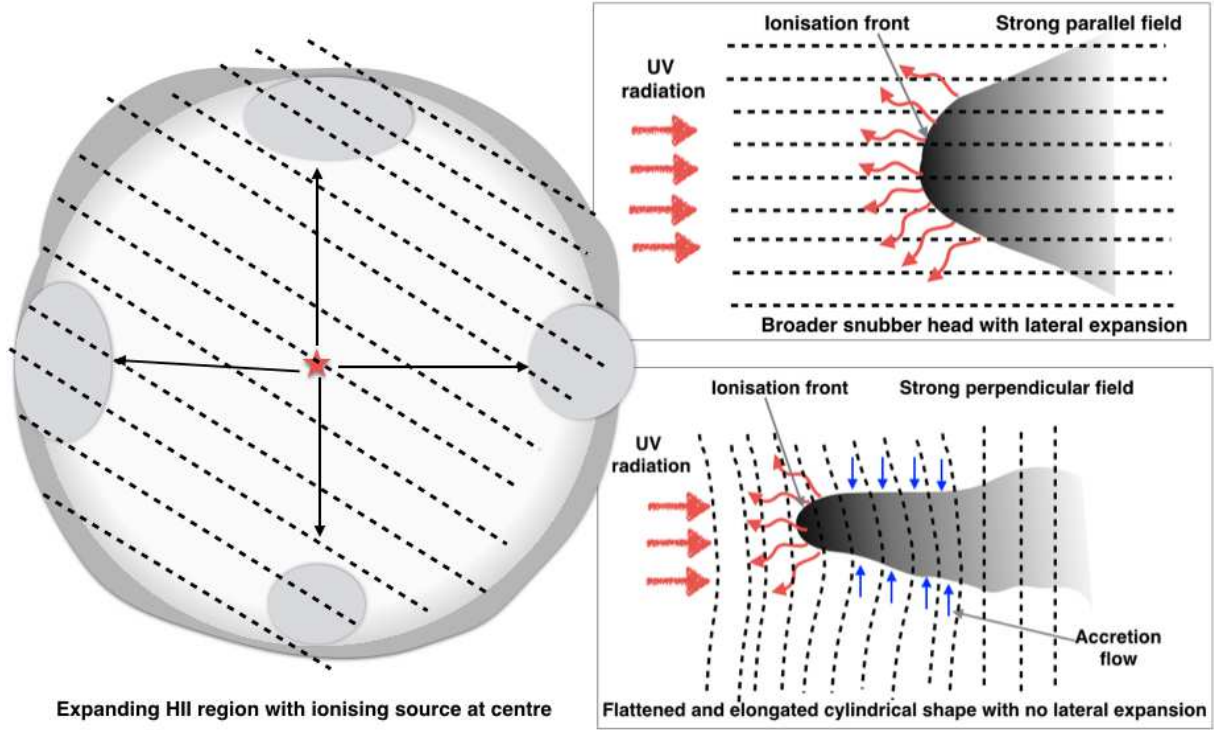}}
\caption{Schematic diagram showing the formation and expansion of H$~$II region due to UV radition from central ionizing source. The formation of BRCs are understood by assuming the pre-existing clumps/clouds on the periphery of Str$\rm \ddot{o}$mgren sphere. The fates of molecular clouds in the presennce of parallel and perpendicular B-fields are also shown in the right column.}\label{Fig:cartoon}
\end{figure*}

To understand the magnetized evolution of the BRCs associated to H$~$II region IC1396 in different directions and considering the MHD simulations by \citet{2009MNRAS.398..157H}, we made a cartoon diagram shown in Fig. \ref{Fig:cartoon}. The left panel in the figure shows the H$~$II region formed by gas, dust and ionized material swept by the high energy UV radiation from HD206267. The H$~$II regions interact with the molecular clouds in the vicinity and pushes away the low density material faster than the high density cloud cores. In Fig. \ref{Fig:cartoon}, we showed the preexisting molecular clumps sitting on the edges of the H$~$II region. The low density part of these clumps might have been ionized and pushed away by the expanding H$~$II region giving rise to the bright rimmed structure. The ambient B-fields are shown with the dashed lines. In the right upper and lower panels, we have shown the fate of the two globules when the field lines are parallel and orthogonal to the ionizing radiation. The diagrams in right panels can probably explain the magnetized evolution of BRC\,38 and IC1396A nebuale, respectively. The dynamical and kinematical motions in these BRCs due to RDI caused by UV radiation will be presented in our following work (Soam et al. in prep., Neha et al. under prep.) using various low and high density molecular gas tracers.

IC1396A globule in H$~$II region IC1396 is also known as elephant trunk nebula. Such structures formed by pure RDI effect is composed of dense head at the front part followed by less dense tails \citep{1994A&A...289..559L, 2000MNRAS.315..713K, 2009ApJ...692..382M, 2011ApJ...736..142B, 2012MNRAS.426..203H}. IC1396A is appeared to be consistent to the RDI-triggered star formation scenario in a detailed study by \citet{2014A&A...562A.131S} using $\textit{Herschel}$ and $\textit{Spitzer}$ data. This is a much larger and more massive globule compared to the clouds modeled by \citet{2011ApJ...736..142B}. Hence the self-gravity in IC1396A must play an important role in its dynamical evolution due to its higher mass \citep[$\sim$200$M_{\odot}$; ][]{1995ApJ...447..721P, 1996A&A...309..581W}. \citet{2004A&A...426..535M} found that the ionized boundary layer (IBL) pressure in this globule is relatively higher compared to the molecular pressure which creates an appreciable pressure imbalance in the cloud. This suggests that the cloud may be currently undergoing RDI process causing a triggered star formation in this globule \citep{2005AJ....130..188S}. 

Fig. \ref{Fig:corr_pol} shows the foreground corrected polarization vectors representing the B-field geometry in IC1396A in zoomed-in panel. The offset of $\sim 35^\circ$ between the mean direction of B-field and the direction of ionizing radiation suggests that the B-fields are slanted w.r.t. the direction of the incident photons from HD 206267. In IC1396A, we noticed that the degree of polarization in some of the sources lying in high density region (inference of the high and low densities are based on the WISE 12$\mu$m emission as described in \citet{2017MNRAS.465..559S}) is relatively low. The B-field observations in this cloud are matching with the MHD simulations performed by \citet{2009MNRAS.398..157H} and \citet{2011MNRAS.412.2079M} in the globules with strong initial B-fields slanted to the direction of ionizing radiation. In such cases, the globule acceleration caused by the rocket effect at the later stages (0.12 Myr) is not radially away from the ionizing radiation. According to these simulations, the globule is swiftly flattens along its longer axis when the B-fields are strong and found slanted to the direction of ionizing radiation. In this case, the tail part of the cloud is shadowed and the surrounding material is accreted along the inclined field lines.

The distribution of YSOs ($\rm H{\alpha}$ emission stars) towards the BRCs studied here are shown using the red circles in Fig. \ref{Fig:corr_pol}. The distribution of YSOs correlate with the masses of the clouds suggesting more number of YSOs in more massive cloud which is relatively closer to the ionizing source. The importance of B-fields in the evolution of dense cores and subsequent star formation can be better understood by probing the field geometries using NIR and submm polarimetric techniques. This work is only limited to investigate the B-fields in the low density regions of the cloud.

\citet{2005AJ....130..188S} and \citet{2006AJ....132.2135S}, using deep $\textit{Spitzer}$ photometry, optical photometry and spectroscopy, found a population of $>$ 200 H${\rm \alpha}$ emission stars with a mixed age of $\sim$ 4 Myr old stars in central Trumpler 37 cluster and $\sim$1 Myr old stars on the front arc of IC1396A nebula. In addition to this, some obscured IR-exess stars were found embedded in the cloud. 
\citet{2009ApJ...702.1507M} reported presence of several embedded Class I/II YSOs in this cloud. Twenty-four low mass and high accretion rate T-Tauri stars were identified by \citet{2011MNRAS.415..103B}. By investigating the distribution of these stars with different age and distances from the ionizing source, above studies proposed that the formation of the stellar population located in front of IC1396A nebula, had been triggered by the high energy radiation from HD 206267. This finding has been supported by the presence of hundreds of YSOs concentrated within and surroundings of the nebula revealed by the high sensitivity optical and mid-infrared surveys. IC1396A has a young star $\rm LkH{\alpha}$349c in the head part. The outflows from the star $\rm LkH{\alpha}$349c has created a cavity of $\sim 0.3$ pc in the globule head. This cavity is seen in IRAC 8$\rm \mu$m, $\rm ^{12}$CO (J=1-0) and $\rm ^{13}$CO (J= 1-0) maps by \citet{1989MNRAS.241..495N}, and the extinction maps obtained from the 2MASS data by \citet{2009ApJ...690..683R} suggesting that the globule is being reshaped from inside.

The foreground corrected polarization vector revealing the B-field structure in BRC 37 is shown in Fig. \ref{Fig:corr_pol}. The globule axis is almost aligned (not exactly) with the incident UV radiation. This suggest that the high energy radiation from HD 206267 might have affected the formation and evolution of this cloud \citep{2001MNRAS.327..788W, 2002ApJ...568L.127F, 2006MNRAS.369..143M}. The mean direction of the B-field is aligned with the longer axis of the cloud. The least value of the offset $\sim 10^\circ$ (among the BRCs studied in this work) between the ionizing radiation and the B-field direction suggests that the initial B-fields are nearly parallel to the UV radiation. The curvature in the cloud's rim might have been caused by the radiation. The alignment of B-fields geometry of the cloud longer axis suggests that the B-fields could be dragged away from the ionizing source. The B-fields will eventually be aligned with the direction towards the exciting star i.e. the cloud longer axis if the B-field is frozen into the gas \citep[e.g., ][]{1989ApJ...346..735B}. The structure of the B-fields in BRC 37 globule is similar to the B-field morphology found in M16 pillar by \citet{2007PASJ...59..507S}. 
 
Hydrodynamical simulation towards pillar like structure such as M16 is presented by \citet{2001MNRAS.327..788W}. These simulations suggest that the narrow pillar like structure will often occur in the H$~$II regions with the large-scale inhomogeneities and their evolution may be resulting into the extended sequences of radiation-induced star formation. The mass of shocked molecular gas at the head of the column like structure is interestingly found closer to the Jeans mass \citep{2001MNRAS.327..788W}. BRC 37 is not in pressure equilibrium between the IBL pressure and the molecular pressure \citep{2004A&A...426..535M} and found with a star IRAS 21388+5622 associated with outflows \citep{1991ApJ...370..263S} and age $\sim$0.3 Myr \citep{1990A&A...233..190D} embedded on the tip of the cloud. The formation of this star might have been triggered by the ionizing radiation. The identification of other near-infrared sources \citep{1995ApJ...455L..39S} nearly aligned with this star and the globule axis suggest the sequential star formation. This cloud is also found to be containing a young brown dwarf candidate which is reported as the first observational evidence of triggered brown-dwarf formation in BRCs \citep{2008AJ....135.2323I}.

The B-field geometry inferred from our polarization studies towards BRC 38 is also shown in Fig. \ref{Fig:corr_pol}. This cloud has the structural similarities with IC 63 globule associated to Sh 2-185 discussed in \citet{2017MNRAS.465..559S}. We were not able to observe B-fields in the high density regions of the cloud BRC 38 using optical polarization technique. Thus the polarization map in Fig. \ref{Fig:corr_pol} shows the B-field geometry on the periphery of the cloud. The offset of $\sim 20^\circ$ between the direction of the ionizing UV radiation and the mean direction of the B-field in the cloud suggests a slanted B-field geometry w.r.t. the ionizing radiation. In BRC 38, particularly two features are noticeable. One is that the field lines exhibit two components on the western face of the globule which is looking towards the ionizing radiation. The other is that there is a density enhancement on the tip of the globule \citep[850 $\rm \mu$m SCUBA map;][]{2004A&A...426..535M}. This may be caused by the accumulation of the material due to the ionizing pressure. The mean direction of B-field is almost consistent to the mean B-field direction in IC1396A globule suggesting that the B-fields shown in these two clouds are well connected to the ambient large scale B-fields in the H$~$II region IC1396.  \citet{2009MNRAS.398..157H} and \citet{2011MNRAS.412.2079M} presented the 3D MHD simulations explaining the cases where initial B-fields are inclined (like found in BRC\,38) to the incident ionizing radiation. In those cases, the lateral compression of the neutral globule is opposed by the B-fields and this results into a broader and snubber globule head with much simpler internal structure. These cases will not show any accretion on the globule tail part because thermal pressure of the ionized gas at the sides of the globule will not be sufficient enough to laterally compress the tail’s longitudinal B-fields. The photoevaporation flows are channeled by the B-fields towards the symmetry axis by means of a focusing shock \citep{2009MNRAS.398..157H}. These simulation results are matching with our B-field observations in BRC\,38.

BRC\,38 is found to be undergoing triggered star formation due to the presence of high energy photons from the star HD206267 in the vicinity. IRAS21391+5802 is embedded on the tip of the nebula and associated with extended bipolar outflows \citep{1989ApJ...342L..87S}. \citet{2002ApJ...573..246B}, using millimeter wavelength observations revealed the other three embedded intermediate and low mass sources named as BIMA 1, 2 and 3 in this cloud. The young stars embedded in the cloud and associated with outflows indicate the ongoing star formation in the dense core at the tip of BRC 38.

The B-field geometry towards BRC\,39 is also shown in Fig. \ref{Fig:corr_pol}. The direction of radiation towards this cloud is at $\sim 90^\circ$. There is a clear bending of the B-fields on the head part of this cloud. These results are similar to the curved field geometries found in the studies towards LBN 437 \citep{2013MNRAS.432.1502S} and B335 \citep{2011ApJ...732...97D}. The mean direction of the B-field is slanted w.r.t. the ionizing radiation from HD 206267. There is a distinguishable elongated density enhancement feature on the lower rim of this cloud. Some of the vectors are found parallel to this feature suggesting that the material could be following the B-field lines in this region. By its structural axis orientation and morphology, BRC 39 is an asymmetrical type B cloud, its formation mechanism can be nicely illustrated by the scenario suggested in \citet{2015MNRAS.450.1017K} which indicates that it might have formed from an initially inclined prolate cloud under RDI effect.

\begin{figure*}
\centering
\resizebox{13cm}{6cm}{\includegraphics{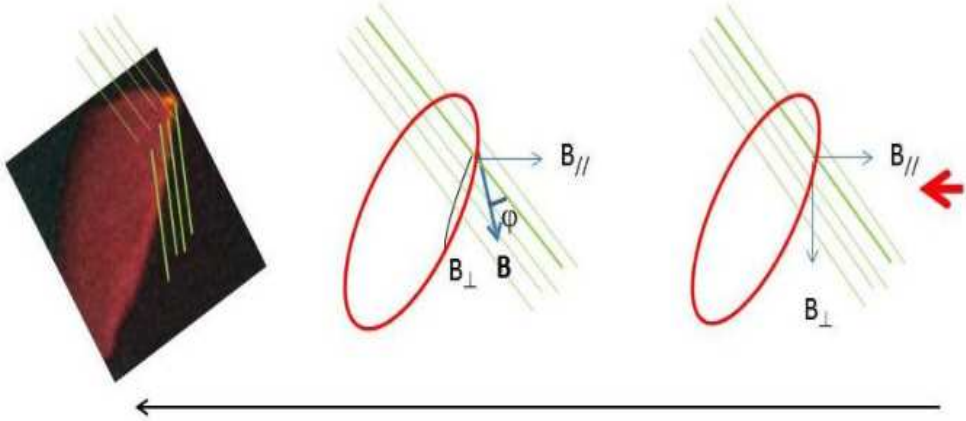}}
\caption{Scematics for depicting structural evolution of BRC\,39 using the simulations from \citet{2015MNRAS.450.1017K}. The evolutionalry track is shown from right to left represented by an arrow. Green segments show the global magnetic field lines and red arrow indicated the direction of ionizing radiation towards the cloud. Two orthogonal components of plane of the sky magnetic fields are labelled as $\rm B_{\parallel}$ and $\rm B_{\perp}$.}\label{Fig:model_BRC39}
\end{figure*}

Fig. \ref{Fig:model_BRC39} shows the schematics adopted from the simulation of \citet{2015MNRAS.450.1017K} to understand the modification in magnetic fields as found in BRC\,39. The right most panel of the figure is a prolate cloud with its structure axis (major axis) inclined to the ionization radiation flux (shown by the red arrow). The global magnetic fields are shown using parallel green lines. After the cloud undergoes RDI process  induced by ionizing radiation, the star-facing surface of the prolate cloud got shocked, and formed an elongated density enhancement thin layer with a highly condensed apex point at the top end, as shown in the rendered density image in the left panel of the figure. This case seems very similar to the observed features towards lower rim of BRC\,39 as shown in the Fig. \ref{Fig:corr_pol}. 

In Fig. \ref{Fig:model_BRC39}, it is seen that the rear side of the cloud is not directly affected by the ionizing radiation and hence there is no significant change from its initial morphology therefore magnetic field there (green lines shown on upper part of the cloud) remains the original direction. On the other hand we can quantitatively analyse the effect on the magnetic fields on the ionizing star-facing side of the cloud (green lines shown on lower part of the cloud). In the right panel, the prolate cloud is initially embedded in the ambient magnetic field of the HII region. One of the green lines (the thicker one) can be taken as an example to show how it decomposes into two components, $\rm B_{\parallel}$ and $\rm B_{\perp}$ to the ionization radiation coming from the direction denoted by the red arrow. After the RDI process, $\rm B_{\parallel}$ will not change but the previously perpendicular component $\rm B_{\perp}$ is curved into the shocked surface layer on the star-facing side (as described by the lower right panel of Fig. \ref{Fig:cartoon}), as shown by the blue curve in the middle panel of the figure. The total magnetic field {\bf B} is now approximately represented by the blue arrowed thick line in the middle panel, with an angle $\rm \varphi$ from the original direction (thick green line). The final configuration of the magnetic field as seen in the lower part of BRC\,39, is depicted by the parallel green lines as shown in the left panel of the figure which is bent from the direction of the original magnetic fields seen in upper part of the cloud.

BRC\, 39 is also found to be the cloud with ongoing star formation activity. The presence of embedded source IRAS 21445+5712 is revealed with a far-infrared luminosity $\sim$96$\rm L_{\odot}$ \citep{2014MNRAS.443.1614P}. Several $\rm H{\alpha}$ emission stars associated to this cloud are reported by \citet{Oguraetal2002} and \citet{2012AJ....143...61N}. The ongoing star formation activity is also suugested by the presence of water maser emission \citep{2005A&A...443..535V} and two bright bow shocks HH 865A and HH 865B emerging from IRAS 21445+5712 \citep{2005A&A...432..575F}

\subsection{Magnetic field strength}\label{MF_strength}

To compare the observation in this work and for understanding the role of B-fields in shaping the clouds in IC1396, we must know the field strength in the region. Fortunately, there are few prolific techniques available now to measure the field strength. One classical method to measure the strength of the B-field projected on to the plane of the sky was proposed by \citet{1953ApJ...118..113C}. This technique was lately modified and new analysis methods are put forth to study the B-field and its interplay with turbulence \citep{2001ApJ...546..980O, 2008ApJ...679..537F}. Using present observations, we estimated the strength of the B-fields using classical Chandersekhar \& Fermi formulation \citep{1953ApJ...118..113C}. This formula requires the information of volume density ($\rm n_{H_{2}}$) in particles per cubic centimeter, the molecular line width in terms of FWHM ($\rm \Delta$V) in $\rm km s^{-1}$ and the dispersion in the measured polarization angle ($\rm \delta\theta$) in degree. The modified form of the formulations is given by \citet{2004ApJ...600..279C} which is

\begin{equation}\label{CF}
{\rm B\textsubscript{pos} = 9.3\sqrt{n_{H_{2}}}\frac{\Delta V}{\delta\theta} \hspace{0.4cm}\mu\,G}
\end{equation}

The equipartition of kinetic and perturbed magnetic energies was assumed while deriving the original equation. Considering the polarization angle deviation and the velocity dispersion in the line of sight, the strength of the B-fields projected on the sky could be estimated. 

The deviation in $\rm \theta$ obtained by fitting Gaussian to the distribution of polarization angles is used to calculate the $\rm \delta\theta$ in the clouds studied here. The dispersion in $\rm \theta$ is then corrected by the uncertainty in $\rm \theta$ using the steps given in \citet{2001ApJ...561..864L} and \citet{2010ApJ...723..146F}. In this procedure, the dispersion is corrected in quadrature by the polarization angle using ${\rm \Delta\theta = ({\sigma_{std}}^{2} - {\langle\sigma_{\theta}\rangle}^{2})^{1/2}}$, where the mean error $\rm \langle\sigma_{\theta}\rangle$ was calculated from ${\rm \langle\sigma_{\theta}\rangle = \Sigma\sigma_{\theta i}/N}$, here $\rm \sigma_{\theta i}$ is the estimated uncertainty and $\rm \sigma_{std}$ is the standard deviation in position angle (for details see \citet{2017MNRAS.464.2403S}). We adopted the CO(J=1-0) line FWHM from our ongoing molecular line survey of BRCs using Taeduk Radio Astronomical Observatory (TRAO; Soam et al. in prep.). We checked the perfect Gaussian profile of CO lines and estimated the average FWHM as $\rm 2.1\pm 0.61~km~s^{-1}$ and $\rm 1.51 \pm 0.07~km~s^{-1}$ towards IC1396A and BRC\,37, respectively. CO(1-0) observations towards SFO\,38 were affected by the outflow activity from the embedded protostar hence we adopted the $\rm C^{18}O$(1-0) line width which is found to be $\rm 1.31 \pm 0.10~km~s^{-1}$ (Neha et al. under prep.). The line width of $\rm ^{13}CO$(2-1) towards BRC\,39 as $\rm1.7~km~s^{-1}$ (due to unavailability of CO data) has been adopted from \citet{1996A&A...309..581W}. The volume densities toward these clouds are calculated using the hydrogen column density information from \citet{2010MNRAS.408..157M} which is found to be $\rm 16\times 10^{21} cm^{-2}$ towards BRC\,39. For other two clouds IC\,1396A and BRC\,38, we considered the average column density as $\rm \sim 10^{22} cm^{-2}$ (based on the values given for all other BRCs in \citet{2010MNRAS.408..157M}). This value is found similar to the column densities obtained by $\rm CO$(1-0) and $\rm C^{18}O$(1-0) observations (Soam et al. in prep., Neha et al. in prep.) using the relation given by \citet{1991ApJ...374..540G}). Using the only angular extents of the clouds where polarization observations are made and taking 750 pc as the distance to IC 1396 H$~$II region, we estimated the volume densities of IC\,1396A, BRC\,38 and BRC\,39 as $\rm 9.3\times 10^{3} cm^{-2}$, $\rm 7.4\times 10^{4} cm^{-2}$, and $\rm 3.8\times 10^{4} cm^{-2}$, respectively. Using the volume density and line width information in CF relation, we estimated the magnetic field strengths as $\rm \sim110\mu$G, $\sim220\mu$G, and $\sim150\mu$G towards IC1396A, BRC\,38, and BRC\,39, respectively. In case of BRC\,38, we have used the polarisation values only on the tip of the cloud for estimating the magnetic field strength because we are adopting the $\rm C^{18}O$(1-0), optically thin tracer, line width in the calculation of B-field strength. Rest of the polarization measurements are mostly projected outside the cloud part therefore we did not use these values. B-field strength towards BRC\,37 has not been estimated due to very high dispersion ($> 25^{\degree}$) in polarization angle which breaks one of the assumptions of the CF relation.

\subsection{Pressure budget in BRCs}\label{IBL_prop}

\begin{table*}
\caption{Pressure estimation towards BRCs studied in this work. The values of IBL pressure are adopted from \citet{2004A&A...426..535M}. The temperature information for estimating thermal pressure are obtained from \citet{2014A&A...562A.131S} towards IC1396A. The temeperature information towards BRC\,37, BRC\,38, and BRC\,39 have been taken from \citet{2000AJ....119..323S}. In the table, IBL, thermal, turbulent and magnetic pressures denoted by $\rm P_{IBL}$, $\rm P_{th}$, $\rm P_{turb}$, and $\rm P_{mag}$, repectively. }\label{tab:mogran}
\scriptsize
\begin{tabular}{llllllr}\hline
Object  & Measured Ionizing flux& Predicted ionizing flux& $\rm P_{IBL}$  & $\rm P_{th}$ & $\rm P_{turb}$ & $\rm P_{mag}$ \\
& $\rm \Phi(10^{18}cm^{-2}s^{-1})$ &$\rm \Phi_{P}(10^{18}cm^{-2}s^{-1})$ & $\rm (10^{-10} erg~cm^{-3})$ & $\rm (10^{-10} erg~cm^{-3})$ & $\rm (10^{-10} erg~cm^{-3})$ & $\rm (10^{-10} erg~cm^{-3})$ \\\hline
IC1396A &2.4 &59.6 &  1.0    &  0.05 &  1.7 &  0.6 \\
BRC\,37 &2.0 &10.2 & 15.0  &  0.12 &  2.6 &  -- \\
BRC\,38 &2.8 &13.3 &  11.0  &  0.05 &  1.0 &  21.0 \\
BRC\,39 &1.9 &9.2 &  --      &   0.08 &  2.1&   3.2  \\\hline
\end{tabular}
\end{table*}

To compare and investigate the importance of various pressures in the BRCs studied here, we estimated the thermal ($\rm P_{th}$), turbulent ($\rm P_{turb}$), and magnetic pressures ($\rm P_{mag}$) towards these clouds. The IBL pressure ($\rm P_{IBL}$) values were adopted from \citet{2004A&A...426..535M}\footnote{They carried out a radio, optical and infrared wavelength imaging survey of 44 BRCs using the NRAO/VLA Sky Survey (NVSS), Digitised Sky Survey (DSS) and the Midcourse Space eXperiment (MSX) archived data}. The datasets characterized the physical properties of the IBL of the BRCs. They compared the measured flux ($\rm \Phi$) and predicted ($\rm \Phi_{P}$) ionizing fluxes incident the rims of these clouds. The measured ionizing fluxes are found to be lesser than the predicted ionizing flux based on the spectral type of ionizing source and the distance of cloud from the source. Though the predicted ionizing fluxes can be considered as the upper limit because the loss in photons caused by absorption from intervening material between the star and the cloud has not been considered by \citet{2004A&A...426..535M} also the distance between star and the clouds are projected values. The values of pressures estimated towards BRCs studied here are shown in Table \ref{tab:mogran}. Various pressures in the BRCs studied here, can be compared in the table. Dynamical pressure in the clouds is found to be lower than the external pressure which suggests that the shock driven by the photoionization could propagate into the molecular cloud, causing it to implode and trigger star formation. Magnetic pressures of the clouds are estimated using the B-field strength over the region where optical polarimetry covers the cloud. Magnetic pressure is found to be lower than the external pressure in these BRCs except in BRC\,38. The magnetic pressure in this cloud is estimated to be around two times the external pressure suggesting that fields can support cloud against external pressure.


\subsection{Polarization in YSOs}

The spectral type and $\rm H{\alpha}$ equivalent width (EW) for some of the YSOs observed in this work towards IC1396 are given by \citet{2005AJ....130..188S, 2006AJ....132.2135S}, and \citet{2010ApJ...717.1067C}. The spherically distributed circumstellar dust around a central YSO producing infrared fluxes lead to a very large extinction along the line of sight at optical wavelength. Hence, the infrared emitting dust must be distributed in a flattened disk-like geometry \citep{1972PASP...84..745S}. This can cause the high polarization when viewed from edge-on. \citet{1972ApJ...175..127B} detected the intrinsic polarization in several pre-main sequence stars (PMS) in NGC 2264. The observed infrared excess emission, $\rm H{\alpha}$ emission and the intrinsic polarization of the YSOs suggest the presence of circumstellar disks around these objects. We found a higher polarization and $\rm H{\alpha}$ EW associated to the stars with higher extinction. Some of the YSOs have low polarization even with the high extinction. The may be due to the dependence of polarization on viewing angle (edge on or face on).

\section{Conclusions}\label{Conclusions}

We have made a first systematic study of B-fields towards multiple BRCs in different directions of H$~$II region IC1396 at different distances from the same central ionizing source. The star, HD 206267 with spectral type O6.5V, located at a distance of $\sim$600 pc, is supposed to be the main ionizing source for H$~$II region IC1396. B-field study of the BRCs in east, west, north and south give an idea of the complete geometry of the B-field in this region. To interpret the B-field geometry, we have subtracted the polarization contribution of ISM dust in the line of sight by observing the stars foreground to the clouds. The corrected results are used to interpret the B-field geometries. The orientations of mapped B-fields are found parallel to that of the Galactic plane suggesting a  well connected field structure from larger scale to cloud scale. It seems that the fields lines in the BRCs studied here, were slanted to the direction of ionizing radiation prior to being affected by it. B-fields in BRC\,37 and BRC\,38 seem to be following the structure of the cloud. In IC1396A, may be due to the presence of strong B-fields, the structure of the globule becomes anvilled and elongated in shape. Our results are matched with the MHD simulations conducted towards BRCs to test their magnetized evolution. The fields lines are bent at the head part of BRC\,39. BRC\,38 represents the case of strong B-fields almost parallel to the direction of ionizing radiation. The B-field strengths are estimated as $\rm \sim110\mu$G, $\sim220\mu$G, and $\sim150\mu$G towards IC1396A, BRC\,38, and BRC\,39, respectively. The pressure budget of these BRCs are tested and it is found that the dynamical pressures in the clouds are lower than the external pressure which suggests that shock driven by the photoionization could propagate into the molecular cloud, causing it to implode and trigger star formation. 


\section{ACKNOWLEDGEMENT}
This research has made use of the Simbad database, operated at CDS, Strasbourg, France. We also acknowledge the use of NASA's \textit{SkyView} facility (http://skyview.gsfc.nasa.gov) located at NASA Goddard Space Flight Center. AS thanks KASI for the post-doctoral research fund. CWL  was supported by Basic Science Research Program through the National Research Foundation of Korea (NRF) funded by the Ministry of Education, Science and Technology (NRF-2016R1A2B4012593). A.S. thanks Dr. Piyush Bhardwaj for the help during the observations.

\bibliographystyle{mnras}
\bibliography{IC1396ref}

\label{lastpage}
\end{document}